\documentstyle[psfig,graphicx]{mn}

\def\gapprox{\lower.4ex\hbox{$\;\buildrel >\over{\scriptstyle\sim}\;$}}
\def\lapprox{\lower.4ex\hbox{$\;\buildrel <\over{\scriptstyle\sim}\;$}}

\def\<{\langle}
\def\>{\rangle}

\newcommand{\ms}{\noalign{\vspace{3pt plus2pt minus1pt}}}

\def\be{\begin{equation}}
\def\ee{\end{equation}}
\def\bea{\begin{eqnarray}}
\def\eea{\end{eqnarray}}
\def\nn{\nonumber}

\def\PAo{2\psi_0}
\def\dPAo{2\Delta\psi}
\def\Io{I_{0}}
\def\Iot{k_0}
\def\EllAo{2\chi_0}
\def\dEllAo{2\Delta\chi}
\def\dfIt{\Delta_2}
\def\dI{\Delta_1}

\title{An empirical model for the polarization of pulsar radio emission}

\author[Melrose, Miller, Karastergiou \& Luo]
      {Don Melrose, Andrew Miller, 
      Aris Karastergiou\thanks{present address: IRAM,
300 rue de la Piscine,
38406 St Martin d'Heres,
France}
 \& Qinghuan Luo\\
        School of Physics, University of Sydney, NSW 2006, Australia}

\date{
          --- Received
         in original form  
        }

\pubyear{2005}

\begin{document}

\maketitle

\begin{abstract}
We present an empirical model for single pulses of radio emission from pulsars based on gaussian probability distributions for relevant variables. The radiation at  a specific pulse phase is represented as the superposition of radiation in two (approximately) orthogonally polarized modes (OPMs) from one or more subsources in the emission region of the pulsar. For each subsource, the polarization states are drawn randomly from statistical distributions, with the mean and the variance on the Poincar\'e sphere as free parameters. The intensity of one OPM is chosen from a log-normal distribution, and the intensity of the other OPM is assumed to be partially correlated, with the degree of correlation also chosen from a gaussian distribution. The model is used to construct simulated data described in the same format as real data: distributions of the polarization of pulses on the Poincar\'e sphere and histograms of the intensity and other parameters. We concentrate on the interpretation of data for specific phases of PSR B0329+54 for which the OPMs are not orthogonal, with one well defined and the other spread out around an annulus on the Poincar\'e sphere at some phases. The results support the assumption that the radiation emerges in two OPMs with closely correlated intensities, and that in a statistical fraction of pulses one OPM is invisible.
\end{abstract}

\begin{keywords}
polarization -- pulsars: general -- pulsars: individual (PSR B0329+54)
\end{keywords}

\section{Introduction}

Recent observations of the polarization of single radio pulses from pulsars \cite{ketal01,ketal02,kjv02} have confirmed much earlier observations \cite{em68,cs69,lsg71} that the polarization can vary substantially from pulse to pulse and that the radiation appears to be a mixture of two orthogonally polarized modes (OPMs) \cite{scetal84,ms00,M02,M03a,J04}. The emphasis in the recent observations has been on the polarization at a specific pulse phase \cite{KJK02,KJK03,ES04,E04,J04}, whereas the emphasis in the early literature was on the variation with time within a single pulse. Observations of the Stokes parameters, $I,Q,U,V$, at selected pulse phases provide statistical information on four parameters at each phase, and these are conventionally represented by the intensity, $I$, the degree of polarization, $r$, and a point on the Poincar\'e sphere, describing the state of polarization, represented by the latitude, $2\chi$, and longitude, $2\psi$. For each chosen phase, the polarization in each pulse is represented by a point on the Poincar\'e sphere, and the OPMs correspond to a distribution of points concentrated around the ends of a diagonal through the sphere. Several different projections of the Poincar\'e sphere have been used to present pulsar data: a Mercator projection \cite{lsg71}, a Hammer-Aitoff projection \cite{KJK03}, and a Lambert equal area projection of the hemispheres  \cite{ES04,E04}. The representation of the data on the Poincar\'e sphere contains information on only two of the four Stokes parameters, and to be complete it needs to be complemented with information on the statistics of the intensity and the degree of polarization, and on the correlations between these two parameters and the polarization.  Such information is available, and is usually represented by histograms. 

The pulse to pulse variation in polarization is probably an important clue to understanding the emission process and its polarization characteristics. The interpretation in terms of OPMs is very strongly indicative of an emission process that generates radiation in two orthogonal modes with correlated intensities. The polarization appears to vary randomly about a mean for each OPM, with the ratio of the intensities also varying randomly. In order to identify the physical processes that lead to these variations it is important to describe them statistically.  McKinnon \& Stinebring \shortcite{ms00} proposed a model in which the polarization of each OPM is always 100\%, and the observed polarization is $<100$\% due to it being the sum of contributions from the two OPMs. Johnston \shortcite{J04} fitted data to this model, assuming gaussian variations with a mean and a variance for the relevant variable. Concentrating on the circular polarization, McKinnon \shortcite{M02} proposed a statistical model, that includes noise convolved with the signal, and found that large fluctuations in circular polarization are possible due to the variations in relative intensity of the two modes. McKinnon \shortcite{M03a} developed this statistical model further by including the other Stokes parameters in it, leading to a probability distribution for the polarization on the Poincar\'e sphere, and McKinnon \shortcite{M04} applied this formalism to observed fluctuations in the position angle (PA) of the linear polarization. Edwards \& Stappers \shortcite{ES04} also used a model that involves a probability distribution on the Poincar\'e sphere in interpreting their data on PSR B0329+54. Our model incorporates all these features with the exception of noise, and adds features relating to nonorthogonality of OPMs and the correlation between the intensities of the OPMs.

In the empirical model described here, we assume gaussian probability distributions to describe each relevant variable in the model. The basic model involves the emission being the sum of the two OPMs each of which has its polarization described by a two-dimensional probability distribution on the Poincar\'e sphere. This involves four parameters for each OPM: two to describe the mean (e.g., the mean latitude and longitude on the Poincar\'e sphere) and two to describe the variance, which we assume to be different in two different directions. There is observational evidence that the intensity variations from pulse to pulse are consistent with log-normal statistics \cite{cjd01}, with notable exceptions such as giant pulses. The statistics are log-normal at different phases, although the statistical parameters can vary with pulse phase \cite{cjd04}. In the model we specify that the intensity of one OPM, identified as mode~1, has log-normal statistics. The mean of $\ln I_1$ is set to zero, without loss of generality. The probability distribution then involves one additional parameter, the variance in $\ln I_1$, whose value, $\dI= 0.8$, we choose to match the observed variance in the natural log of the intensity. The interpretation of the data requires that the intensities of the two OPMs be neither strictly correlated or completely uncorrelated: they must be partially correlated. We model the  partial correlation by assuming that the intensity in mode~2 is $k$ times that in mode~1 (in a given pulse) with $k$ described by a gaussian distribution with a mean $\Iot$ and a variance $\dfIt$. A notable result is that we find that some features require $\Iot$ close to but not equal to unity, and $\dfIt$ small, implying a tight correlation.

Noise is not included in our gaussian probability distributions; it was included in probability distributions by McKinnon \shortcite{M02,M03a,M04}. We could include noise by convolving each relevant probability distribution with a noise distribution. However, this would increase the number of free parameters in the model, and complicate the identification of the important parameters needed to explain the features on which we concentrate here. Our neglect of noise implies that the model applies only to data for which instrumental noise and other noise is known to be unimportant. 

To illustrate the use of the model we focus on the interpretation of the data presented by Edwards \& Stappers \shortcite{ES04} for eight different phase bins of PSR B0329+54. For some of these bins the data are well described, at least to a first approximation, by an OPM model with slightly elliptically polarized modes and some random variations in the polarization of each OPM, with the means being orthogonal. For other phase bins the OPMs are not orthogonal, and in several examples one mode is spread out around an annulus on the Poincar\'e sphere. We focus on the data that seem most difficult to explain within the model: nonorthogonality of the OPMs, and asymmetry between the OPMs. 

The probability distributions are introduced in Section~2, and the results of specific simulations are presented in Section ~3. Physical interpretations are discussed in Section ~4, and the conclusions are given in Section ~5.
 
\section{Probability distributions}

The model is based on gaussian probability distributions for each of the relevant variables: these distributions are defined in this section.

\begin{figure}
\begin{center}
\psfig{file=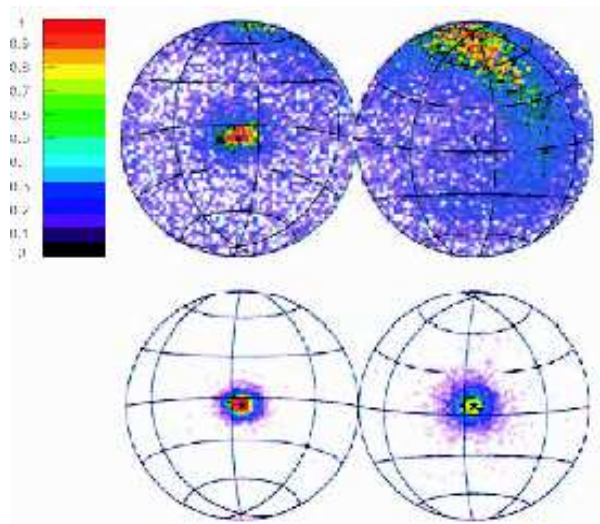,width=8cm}
\caption{The polarization of single pulses of PSR B0329+54 (at $328\rm\,MHz$) are shown for two of eight specific phases in Fig.~2 of Edwards \& Stappers (2004). Each panel contains the Lambert projection of each hemisphere of the Poincar\'e sphere, with the colour code showing the density of points (one point per pulse). All eight lead to significantly different distributions, with only the first and last (shown here in the lower panel) having obvious simple interpretations in terms of OPMs. The example shown in the upper panel has one reasonably well defined mode (left hemisphere) with the other mode seemingly spread out along an annulus on the sphere. The lines in the lower panel are for constant latitude, $2\chi$, and longitude, $2\psi$.
}
\label{figES}
\end{center}
\end{figure}

\subsection{Representation on the Poincar\'e sphere}

A state of polarization is represented by a point on the Poincar\'e sphere. The longitude, $2\psi$, on the sphere determines the position angle of linear polarization, ${\rm PA}=\psi$. Similarly, the latitude, $2\chi$, determines the axial ratio of the polarization ellipse, $T=\tan\chi$: right hand circular, $T=1$, corresponds to the north pole, $2\chi=\pi/2$, and left circular, $T=-1$, to the south pole, $2\chi=-\pi/2$. Points in between the pole and the equator represent elliptical polarizations. In terms of the Stokes parameters, it is convenient to write $q=Q/I$, $u=U/I$, $v=V/I$, so that the degree of polarization is $r=(q^2+u^2+v^2)^{1/2}$. It is only the polarized part that can be represented by a point on the Poincar\'e sphere, and this corresponds to \cite{MMcP91}
\bea
{q\over r}=\cos(2\chi)\cos(2\psi),
\qquad
{u\over r}=\cos(2\chi)\sin(2\psi),
\nn
\\
\ms
{v\over r}=\sin(2\chi).
\qquad\qquad\qquad\qquad
\label{eqn0}
\eea
Information on $I$ and $r$ is not included in this representation, and must be presented in some complementary way.

A specific example of the representation of the polarization of individual pulses on the Poincar\'e sphere is shown in Figure~\ref{figES}, and is discussed further in Section ~3. The number of pulses with polarization close to a given point on the sphere may be represented as a probability density for the polarization represented by that particular point. 

\subsection{Polarization probabilities}

In the model the emission in a single pulse is assumed to be a sum of completely polarized components, consisting of two OPMs for each of $N$ subsources. The polarization state for a particular OPM and particular subsource is drawn from a two-dimensional gaussian distribution on the Poincar\'e sphere. This formalizes a suggestion made by Karastergiou, Johnston \& Kramer \shortcite{KJK03} that, from pulse to pulse, the polarization vectors vary randomly about a mean. The mean polarization is represented by a point, $\chi=\chi_i$, $\psi=\psi_i$ say, with $i=1,2$ labeling the two modes. The mean polarizations may also be described by $q_i, u_i,v_i$, which are related to $\chi_i,\psi_i$ by (\ref{eqn0}) with $r=1$. We assume that the probability distributions around these means may be treated independently for the two modes. The probability distribution around mode~$i$ may be plotted on the Poincar\'e sphere in terms of contours of constant probability. The data are suggestive of distributions that are anisotropic, favoring one direction over another, and to allow for this we assume that the contours of constant probability are ellipses. If these surfaces were circles we would need only one parameter to describe the variance in the polarization for mode~$i$, and with the generalization to ellipses of constant probability we require three parameters to describe the variance for each mode (the extra two being say the axial ratio and the direction of the major axis of the ellipse). To minimize the number of parameters, while still allowing for anisotropy, we adopt the following prescription that fixes the orientation of these probability ellipses.

Let the axes on the Poincar\'e sphere be rotated such that the new north pole coincides with the mean polarization point for mode~$i$, and let quantities relative to the rotated axes be denoted by primes. Thus this choice corresponds to $v'_i=1$, $q'_i=u'_i=0$. If there were no anisotropy, a gaussian probability distribution with variance $a_i^2$ would be proportional to $\exp[-(q'^2+u'^2)/2a_i^2]$. We include an anisotropy and fix its orientation by choosing probability distributions in $q'$ and $u'$ with different variances, $a_i^2$ and $b_i^2$, respectively. With this prescription, the $q'$ are drawn randomly from the distribution
\be
p_{q'}={e^{-q'^2/2a_i^2}\over(2\pi a_i^2)^{1/2}},
\qquad -1<q'<1
\label{eq1}
\ee
and the $u'$ are drawn randomly from 
\be
p_{u'}={e^{-u'^2/2b_i^2}\over(2\pi b_i^2)^{1/2}},
\qquad -1<u'<1.
\label{eq2}
\ee
Then one has
\be
v'=(1-q'^2-u'^2)^{1/2}.
\label{eq3}
\ee
The two-dimensional distribution on the sphere is then represented by expressing the product of the probabilities (\ref{eq1}) and  (\ref{eq2}). The ratio of the semi-axes of the ellipses of constant probability is equal to $a_i/b_i$. On rotating back to the original axes, this determines the orientation of probability ellipses in terms of the original (unprimed) variables centered on the mean polarization state ($2\chi_i$, $2\psi_i$). For linearly polarized modes ($q_i=\pm1$, $u_i=v_i=0$), these ellipses are oriented along latitude and longitude on the Poincar\'e sphere. In the cases discussed below the mean polarization is nearly linear, and these probability ellipses are then oriented nearly, but not exactly, along latitude and longitude.

\subsection{Nonorthogonal OPMs}

Strictly orthogonal OPMs correspond to antipodean points on the Poincar\'e sphere: their PAs differ by $90^\circ$ and their degrees of circular polarization are equal in magnitude and opposite in sign. This corresponds to $2\chi_2=-2\chi_1$, $2\psi_2=2\psi_1+\pi$. Nonorthogonal OPMs correspond to nonzero differences $\dEllAo=2(\chi_2+\chi_1)$, $\dPAo=2(\psi_2-\psi_1)-\pi$. These two differences are free parameters in the model.

\subsection{Intensities of the two modes}

One mode, mode~1 say, is assumed to have a log-normal intensity distribution: 
\be
p_{I_1}={e^{-(\ln I_1-\ln I_0)^2/2\Delta_1^2}\over(2\pi\Delta_1^2)^{1/2}},
\label{eq4}
\ee
where $\ln I_0$ and $\Delta_1$ are the mean and the variance of $\ln I_1$, respectively. The  intensity of mode~2 is chosen to be proportional to  the intensity of mode~1, with the constant of proportionality, $k$, selected randomly from a gaussian distribution with mean $\Iot$ and variance $\dfIt$: 
\be
I_2=kI_1,
\qquad
p_k= {e^{- (k -\Iot)^2/2\dfIt^2}\over(2\pi\dfIt^2)^{1/2}}.
\label{eq4a}
\ee
This ensures that the intensities of the two modes are partially correlated, as suggested by observations \cite{scetal84,KJK02}. 

\subsection{Number of subsources}

One interpretation of the microstructure in pulses is that at any given instant the observer received radiation from a number of separate subsources \cite{L04}. Also, using a carousel model to explain drifting subsources, Edwards, Stappers  \& van Leeuwen \shortcite{ESvL03} argued that some data suggest that two subsources might be visible simultaneously. Our empirical model allows us to regard each pulse as composed of emission from $N$ separate subsources, with different choices of parameters for each subsource. To avoid introducing too many free parameters, in the simulations reported here, when there is more than one subsource, all the subsources have the same probability distributions. Each subsource then corresponds to an independent set of random numbers chosen according to the specified probability distributions. Even in this simple case, the results for $N$ subsources are not related as simply to that for a single subsource as might be anticipated.

\subsection{Additional parameters}

A specific difficulty arises in explaining phases where one of the OPMs is concentrated around a well-defined mean polarization and the other is spread out around an annulus on the Poincar\'e sphere. An additional assumption seems to be required to account for these two features being present simultaneously. We allow this by introducing an additional parameter which is a probability that mode~2 is present and mode~1 is invisible. A physical effect that allows this is that the ray for mode~2 from a given subsource reaches the observer, but the ray for mode~1 from that subsource does not because it is refracted into a different direction \cite{AM82}. For each pulse, the model randomly decides whether to add the orthogonal modes from all the subsources or only a fraction of them. This adds a new parameter  to the model: $p_v$, the fraction of pulses for which mode~1 is invisible. For multiple subsources, $N>1$, we assume that $p_v$ applies only to a fraction, $p_N$, of the subsources.

\section{Results of simulations}

We present six examples of our simulations. The parameters chosen for these six cases are summarized in Table~1. Others parameters not listed, $P =15000$ (the number of pulses in each simulation), $\Io= 1.0$, $\dI= 0.8$, $\PAo=90^\circ$, $\EllAo= -5^\circ$ and $a_1=0.05$ are the same for all the simulations reported here. In presenting the results of specific simulations, we start with the case of a single subsource with orthogonal OPMs, then relax the assumption of orthogonality and the assumption of a single subsource. To focus our simulations we attempt to identify parameters such that our results simulate the specific observations shown in Figure~\ref{figES}. 

\begin{figure}
\begin{center}
\psfig{file=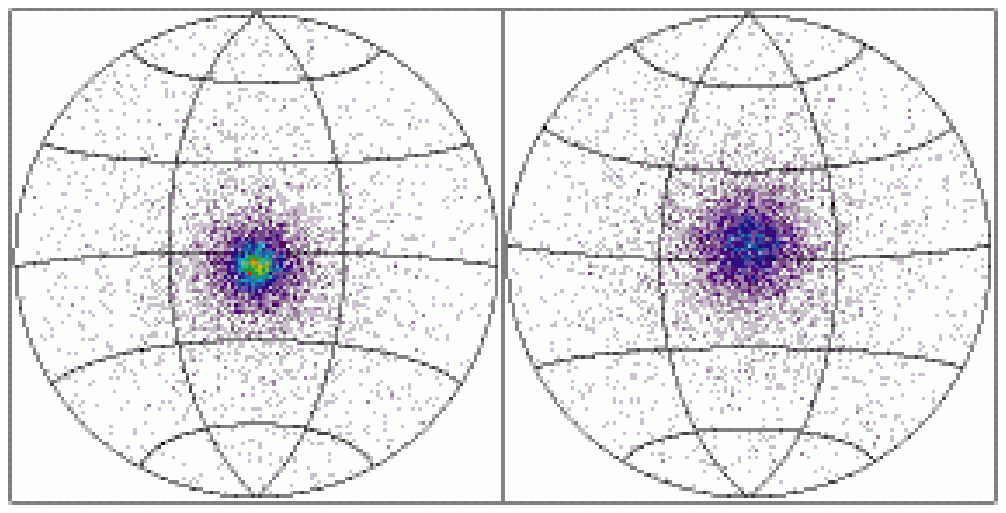,width=8cm}
\vspace{24pt}
\psfig{file=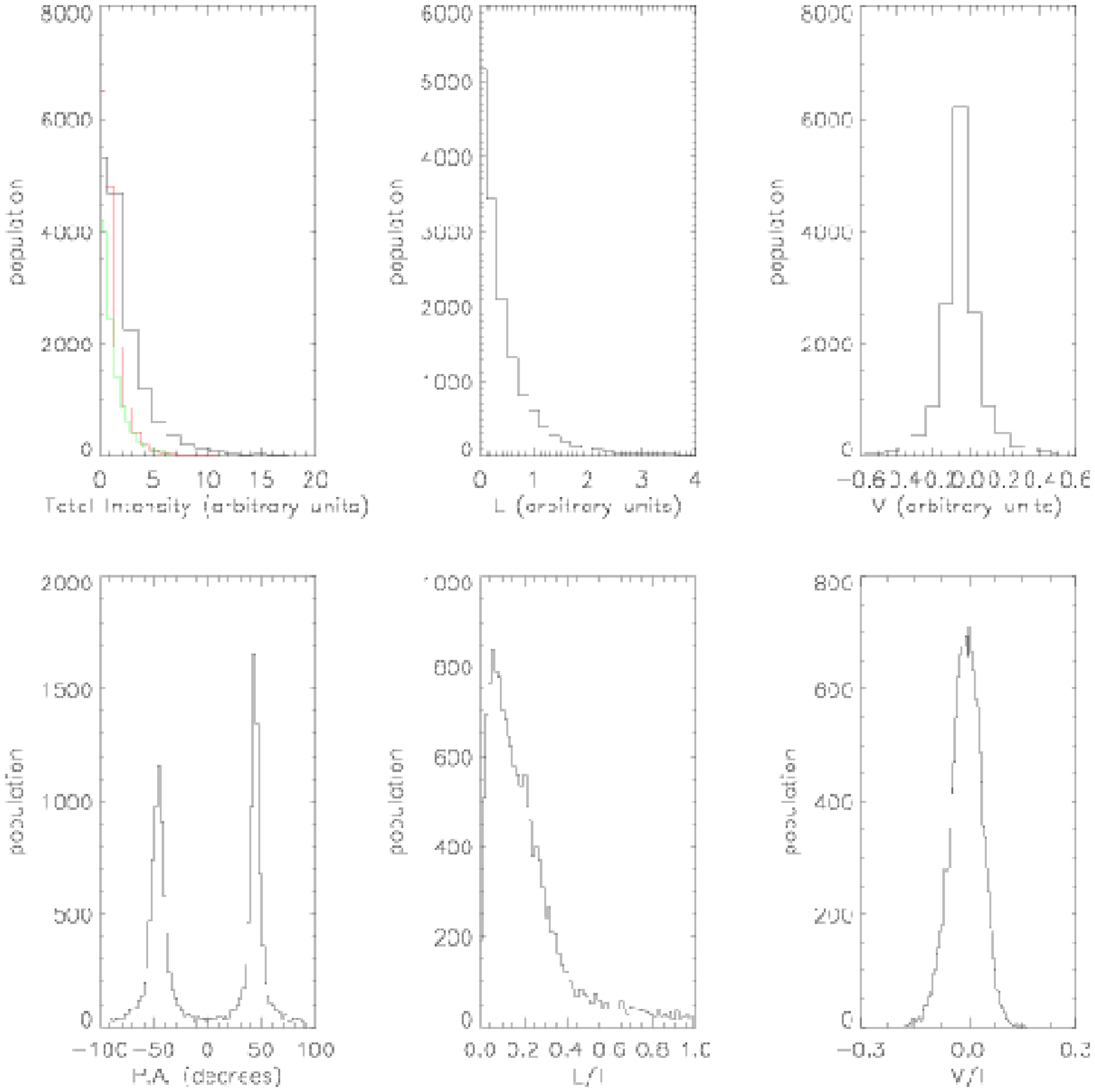,width=8cm}
\caption{An example of the output of the model is shown. The top panel shows the polarization on the Poincar\'e sphere in the same format as Figure~\ref{figES}. The polarization points in the right hemisphere are dominated by mode~1 and those in the left hemisphere by mode~2. The first three histograms show the intensity summed over the set of simulated pulses: the first panel shows the total intensity (in the color version the total intensity is in black and its components in two OPMs are in red and green), and the second and third panels show the linearly and circularly polarized intensities, respectively. The fourth panel shows the PA of the linear polarization, and the final two panels show the degrees of linear and circular polarization. In this example the OPMs are orthogonal and their variances are the same and independent of direction on the Poincar\'e sphere. There is only one subsource. The ratio of the intensity of mode~2 to mode~1 has a mean of unity and a variance of $0.5$. The other specific parameters are given in the text.
}
\label{fig1}
\end{center}
\end{figure}

\begin{table*}
 \centering
 \begin{minipage}{140mm}
  \caption{The parameters used in the simulations shown in Figures~2 to~7 are tabulated. Angles are in degrees. The parameters $P=15000$,  $\Io= 1.0$, $\dI= 0.8$, $\PAo=90^\circ$, $\EllAo= -5^\circ$ and $a_1=0.05$ are the same in all figures, and the parameter $p_N$ is not applicable (NA) for a single subsource, $N=1$. }
  \begin{tabular}{ccccccccccc}
   Fig. &$b_1$
          &$\dPAo$  & $\dEllAo$ & $a_2$& $b_2$&$\Iot$
        &$\dfIt$&$N$&$p_v$& $p_N$\\[6pt]
2&0.05&0&0&0.05&0.05&1.0&0.5&1&0&NA\\
3&0.15&0&0&0.02&0.01&1.1&0.5&1&0&NA\\
4&0.15&0&0&0.02&0.01&1.1&0.05&1&0&NA\\
5&0.15&3&10&0.02&0.01&1.1&0.05&1&0&NA\\
6&0.15&3&10&0.02&0.01&1.1&0.05&1&0.2&NA\\
7&0.15&3&10&0.02&0.01&1.1&0.05&3&0.2&0.4
\end{tabular}
\end{minipage}
\end{table*}

\subsection{Method of calculation}

In attempting to find a model that simulates the data shown in Figure~\ref{figES}, we start by making the simplest assumptions of a single subsource ($N=1$) with orthogonal OPMs ($\dPAo= 0$, $\dEllAo= 0$). With this first example, the polarizations of the two modes are orthogonal, with equal variances that are independent of orientation on the Poincar\'e spheres, and with the same mean intensity. The simulated data are shown in Figure~\ref{fig1}. The first panel is a Lambert projection of the two hemispheres of the Poincar\'e sphere, in the same format as used in Figure~\ref{figES}. The hemispheres are roughly centered on the two mean polarizations, mode~1 in the right hemisphere and mode~2 in the left hemisphere. The information on the intensity is shown in the first of the six histograms, with the intensities in the two modes shown separately in red (mode~1) and green (mode~2). The second and third and fourth histograms show the linearly polarized intensity ($L=(Q^2+U^2)^{1/2}$), the circularly polarized intensity and the distribution in position angle. The usefulness of these histograms is for comparison with data, which can be represented as histograms of $I,L,V$ and PA with minimal processing. The final two panels show the degree of linear and circular polarizations. We construct such histograms for all our examples, but here we show only the Poincar\'e sphere for most of them.

The simulation shown in Figure~\ref{fig1} reproduces the example shown in the lower panel in Figure~\ref{figES} reasonably well. It is clear that this and similar looking examples may be simulated in terms of two orthogonal modes with random spreads about the two polarizations. However, it is also clear that additional assumptions need to be made to simulate the example shown in the upper panel in Figure~\ref{figES}. We attempt to do so by relaxing one-by-one some of the assumptions made in the simulation shown in Figure~\ref{fig1}. Also note that despite the variances in the polarizations being the same, and the mean intensities being the same, for both modes, the distributions on the Poincar\'e sphere are different; this difference is particularly notable in the histogram for PA. These differences are due entirely to the different statistics for the intensities of the two modes: in this case the highest intensities are dominated by mode~2, with $\langle I_2\rangle=\langle I_1\rangle$ and $\langle I_2^2\rangle=1.25\langle I_1^2\rangle$.

\begin{figure}
\begin{center}
\psfig{file=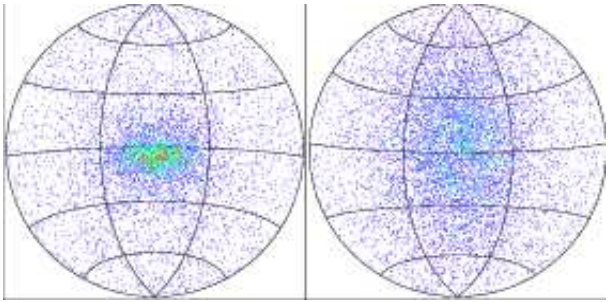,width=8cm}
\caption{Similar to Figure~\ref{fig1} with only the Poincar\'e sphere shown; the centroids of the OPMs remain orthogonal but the variances are different in different directions on the sphere and different for the two OPMs (details in the text). 
}
\label{fig2}
\end{center}
\end{figure}

\subsection{Orthogonal OPMs, different variances}

\begin{figure}
\begin{center}
\psfig{file=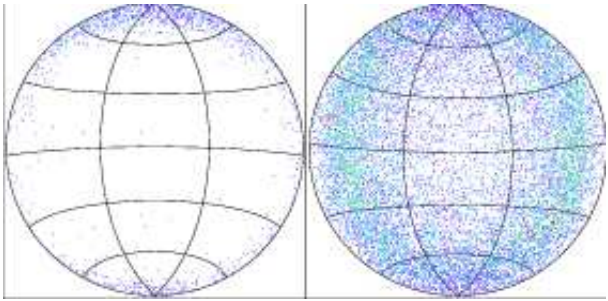,width=8cm}
\caption{As for Figure~\ref{fig2} but with the variance in the ratio of the intensity of mode~2 to mode~1 reduced from $\dfIt=0.5$ to $\dfIt=0.05$ so that there is a strong tendency for the mean polarizations to cancel.}
\label{fig3}
\end{center}
\end{figure}

The example illustrated in Figure~\ref{fig2} differs from that in Figure~\ref{fig1} in that the variances in the polarizations of the two modes are different, and the mean of the ratio of the intensities of the two modes is increased to slightly greater than unity ($\Iot=1.1$). For mode~1, the choice
$a_1=0.05$,
$b_1=0.15$,
implies that the spread is larger on the Poincar\'e sphere in the vertical than the horizontal direction; the surfaces of constant probability are ellipses with axial ratio $b_1/a_1=3$, favoring circular over linear polarization. For mode~2, the choice
$a_2=0.02$,
$b_2=0.01$,
corresponds to a smaller spread than for mode~1, and such that the probability ellipses have axial ratio $b_2/a_2=1/2$, favoring linear over circular polarization. The effect, shown in Figure~\ref{fig2}, is to spread out the points around the mean for mode~1 (on the right) much more strongly than for mode~2 (on the left). It is clear that while such spreading might be one ingredient in attempting to simulate the upper panel in Figure~\ref{figES}, it cannot explain the concentration of the points around a broad annulus, rather than a central modal point. The increase in $\Iot$ from unity in Figure~\ref{fig1} to $\Iot=1.1$ in Figure~\ref{fig2} is relatively unimportant; this ratio becomes more important in Figures~\ref{fig3} to~\ref{fig6} where we attempt to simulate the annulus in Figure~\ref{figES}.

The example illustrated in Figure~\ref{fig3} differs from that in Figure~\ref{fig2} in that the spread in the ratio of intensities is much smaller: the parameter $\dfIt=0.5$ in Figure~\ref{fig2} is replaced by $\dfIt=0.05$ in Figure~\ref{fig3}. This causes the distribution of points to become even more strongly spread out, and for the concentration of points around the means to disappear. In interpreting this, first consider a case (not shown) where there is no spread in the ratio of the intensities of mode~2 to that in mode~1, $\Iot=1$, $\dfIt\to0$. Then, on summing over the Stokes parameters for the two modes the mean polarizations cancel. The cancellation is not exact because the parameters for each mode correspond to two different choices of random numbers. The net polarization is determined by the difference between the two modes, and the degree of polarization is necessarily small, $[(Q_1+Q_2)^2+(U_1+U_2)^2+(V_1+V_2)^2]^{1/2}\ll I_1+I_2=2I_1$, due to $(Q_2,U_2,V_2)\approx-Q_1,U_1,V_1$. This leads to a broad distribution of polarization on the Poincar\'e sphere. The difference between Figure~\ref{fig2}, $\dfIt=0.5$ and Figure~\ref{fig3}, with $\dfIt=0.05$ is that the latter is effectively indistinguishable from the case $\dfIt\to0$ in which the mean polarizations cancel exactly. 

This example adds a further ingredient that is plausibly needed in the interpretation of the upper panel in Figure~\ref{figES}: spreading out of the points due to near equality of the intensities in the two modes. However, the associated loss of concentration around the mean polarizations for the two modes is not consistent with the observations, and a further assumption is needed to overcome this. Before considering how this can be achieved, we relax the assumption of orthogonality.

\subsection{Nonorthogonal polarizations}

\begin{figure}
\begin{center}
\psfig{file=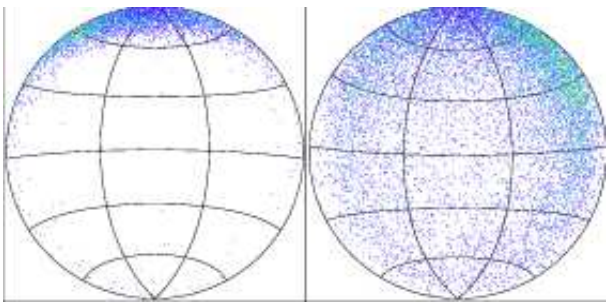,width=8cm}
\caption{As in Figure~\ref{fig3} except that mode~2 is not orthogonal to mode~1, with the difference being
$\dPAo=3^\circ$,
$\dEllAo=10^\circ$.
}
\label{fig4}
\end{center}
\end{figure}

The example shown in Figure~\ref{fig4} differs from Figure~\ref{fig3} only in that the centroid for mode~2 is not orthogonal to that for mode~1, and is displaced from the antipodal point by $\dPAo=3^\circ$,
$\dEllAo=10^\circ$. This introduces a favored direction on the Poincar\'e sphere, and the polarization points tend to form an annulus around this preferred direction.  This annulus provides a natural explanation for the spreading apparent in the observational example in the upper panel of Figure~\ref{figES}. However, the absence of a noticeable concentration of points around the mean polarization for mode~2 is inconsistent with the observations.

Note that the annulus appears when the ratio of the mean intensities is close to unity, but slightly greater than unity ($\Iot=1.1$ here), and the variance ($\dfIt=0.05$ here) corresponds to a standard deviation roughly equal to the difference ($\dfIt\sim(\Iot-1)^2$). The large spread is due to the small variance in this ratio causing there to be a large number of weakly polarized bursts due to the mean polarizations nearly canceling. 

\subsection{Sometime invisible mode}

There is an obvious way of rectifying the absence of concentrations of points around the modal point in Figures~\ref{fig3} and~\ref{fig4}: include a component of the emission that has such a concentration. One way in which this can be achieved in our model is to introduce one or more additional subsources that have such concentrations. However, it is of interest to explore an alternative way of achieving the same result with only one subsource: assume that some pulses are visible in mode~2 but not in mode~1. This is the purpose of our parameter $p_v$ which specifies the fraction of pulses in which mode~1 is invisible.

In Figure~\ref{fig5} we show how Figure~\ref{fig4} is modified by introducing the parameter $p_v=0.2$, which corresponds to mode~1 being invisible in 20\% of the pulses. This adds the peak around the centroid for mode~2 in  Figure~\ref{fig5}, compared with Figure~\ref{fig4}. This peak is due entirely to the pulses for which the component in mode~1 is assumed to be invisible. In both Figure~\ref{fig5} and Figure~\ref{fig6}, there are three peaks in the histogram for PA, at $\approx -90^\circ$, $\approx -10^\circ$ and $45^\circ$. The first of the three can be seen to wrap around at $+90^\circ$, due to the PA being modulo $180^\circ$. The peak at $45^\circ$ corresponds to the cases where mode~1 is invisible. As for the two other peaks at negative PA, the picture is similar to what one expects for nonorthogonal modes \cite{M03b}: they are separated by $<90^\circ$ with a bridge between them. This is due to the addition of the non-orthogonal polarizations of modes~1 and~2, as described for Figure~\ref{fig4}. In fact a similar feature appears in the histogram for Figure~\ref{fig4}, which is not shown. Besides the nonorthogonality of the centroids of modes~1 and~2, the different probability distributions for the polarizations of the two modes is an important ingredient in the structure of the PA histogram.

The annulus first seen on the Poincar\'e sphere of Figure~\ref{fig3}, remains in Figures~\ref{fig5} and~\ref{fig6}. As discussed above, this is due to the almost constant ratio of intensities and the different
shapes of the distributions of modes~1 and~2, which results in instantaneous nonorthogonal polarizations being added.

\subsection{Multiple subsources}

\begin{figure}
\begin{center}
\psfig{file=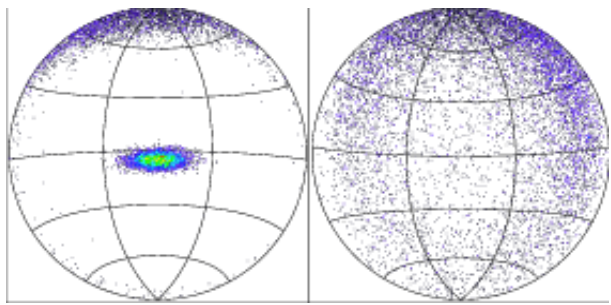,width=8cm}
\vspace{24pt}
\psfig{file=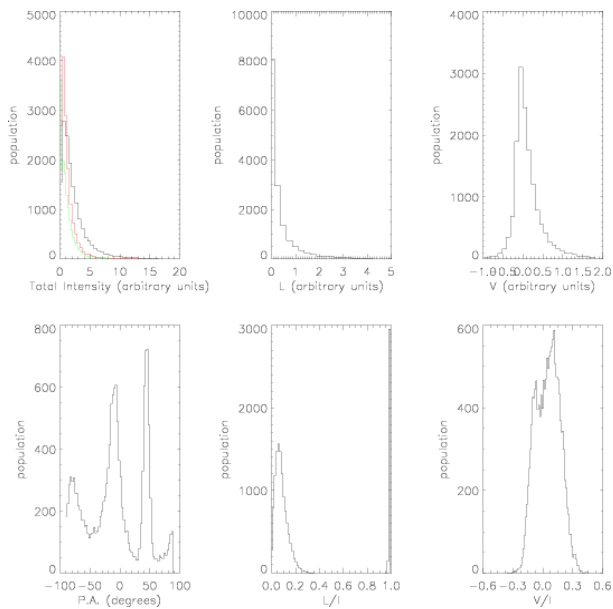,width=8cm}
\caption{As for Figure~\ref{fig4} but with an a fraction $p_v=0.2$ of the pulses having mode~1 invisible.
}
\label{fig5}
\end{center}
\end{figure}

\begin{figure}
\begin{center}
\psfig{file=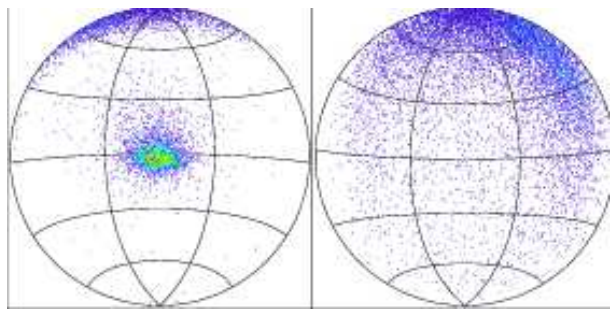,width=8cm}
\vspace{24pt}
\psfig{file=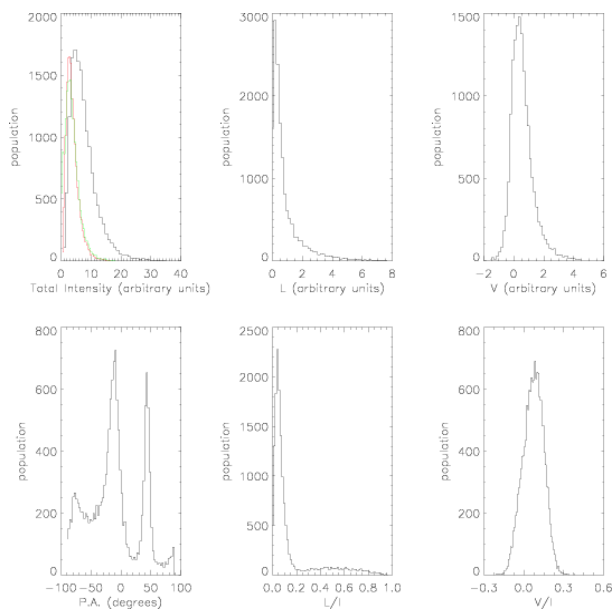,width=8cm}
\caption{As for Figure~\ref{fig5} but with $N=3$ subsources, The fraction $p_v=0.2$ of pulses with mode~1 invisible does not apply to a fraction $p_N=0.4$ of the subsources, which have $p_v=0$.
}
\label{fig6}
\end{center}
\end{figure}

The introduction of the parameter $p_v$ has another consequence that is evident in the fifth histogram in Figure~\ref{fig5}: there is a narrow peak in the degree of linear polarization, corresponding to a small fraction of 100\% polarized bursts. This is an artefact of allowing a fraction of the pulses to be completely polarized in mode~1. This artefact can be eliminated by allowing multiple subsources. An example is shown in Figure~\ref{fig6}, which differs from Figure~\ref{fig5} in that there are $N=3$ subsources. In this case the parameter $p_v=0.2$ is complemented with $p_N=0.4$. 

The distribution of points in Figure~\ref{fig6} is similar to that in the upper panel in Figure~\ref{figES}, and we could improve the agreement by further adjustment of parameters. However, it is also clear that there is no unique way of simulating this particular example. We chose to introduce the parameters  $p_v$ and $p_N$ to overcome the loss of concentration around the centroid of mode~2 when modeling the spread distribution for mode~1 with only one subsource. There is a wide degree of freedom if one assumes that the different subsources can have significantly different properties. One could postulate different subsources to produce different features in any given observation; specifically, one subsource to produce the concentration of points for mode~2, with a negligible contribution from mode~1 ($\Iot\gg1$) and a second subsource to produce mode~1 spread out around an annulus, with no significant focus for mode~2, as in Figure~\ref{fig4}.

\section{Physical interpretation}

In discussing the physical ingredients in our model we start with some general comments, then concentrate on two features of the observations that seem difficult to explain: nonorthogonality of the OPMs, and asymmetry between the OPMs.  We also comment on the difficulty of explaining the separation into two modes with comparable intensities and the limiting polarization.

\subsection{Physical ingredients}

Important physical ingredients in understanding the polarization of pulsar radio emission include birefringence, gyrotropy, ray tracing, separation into two natural modes and limiting polarization. The first three of these ingredients are relatively well understood in principle, although there remain major uncertainties in applying them to pulsars. Birefringence is present in any magnetized plasma: there are two natural wave modes in the medium, and these have different refractive indices and different polarizations. Gyrotropy implies that the natural modes are elliptically (rather than linearly) polarized, and in the present context gyrotropy is due to differences between the distributions of electrons and positrons \cite{ML04}. Under most conditions the natural modes are orthogonally polarized, which implies that their polarization ellipses are orthogonal with opposite handedness. If the medium is inhomogeneous, then birefringence implies that the rays corresponding to the two natural modes propagate along different ray paths, and ray tracing involves determining these paths for specific models of the inhomogeneous plasma \cite{ba86,P01,P02}. 

\subsection{Nonorthogonality and asymmetry between OPMs}

Observationally, there are clear examples where the observed polarizations are concentrated around two mean polarizations that are not orthogonal to each other. However, we do not regard the nonorthogonality of OPMs as implying a nonorthogonality of the natural modes of the medium. OPMs refer to the polarizations seen by the observer. Tracing the rays for each OPM back to the source, the birefringence implies that these originated from different points in the source, with slightly different angles of emission, such that the rays are parallel and coincident at infinity. Physically, the nonorthogonality of the OPMs reflects the difference in the polarizations of the modes at these points and with these directions of emission, and does not reflect an actual nonorthogonality of the natural modes at any given point in the medium. An estimate of the nonorthogonality may then be used to infer a constraint on the properties of the emission given a model for the medium through which the radiation propagates to the observer.

Nonorthogonality of the modes is included in two ways in the model. Besides the explicit nonorthogonality in the centroids of the modes, there is also a nonorthogonality arising from the choice of random number from the independent distributions of the polarizations for the two modes in each simulated pulse. We envisage the source being composed of many subsources, whose polarizations differ due to small differences in relevant parameters between them. Pulse-to-pulse variations in both the source regions and along the ray paths for the two modes are assumed to lead to a degree of randomness in the observed polarization that is described by both the centroids for the two modes being different and the variations about these centroids being independent for the two modes.

Consider the interpretation of the upper panel in Figure~\ref{figES}, which is an example where the two modes are asymmetric, with one concentrated around a centroid, and the other spread out, with no noticeable concentration around the antipodal point. Our model shows that this is consistent with two OPMs provided that (i) the OPMs are not strictly orthogonal, and (ii) there are two subclasses of pulses with (a) the intensities nearly equal in one subclass, leading to the spread out distribution, and (b) one mode is invisible in the other subclass, leading to the concentrated distribution of points. Although some nonorthogonality in the  means of the modes is required, it is also clear that there are two other important ingredients in simulating this case: the spread in polarization about the means, and a tight (but not exact) correlation of the intensities of the two modes (in the first of the two subclasses of pulses). Physically, this emphasizes the importance of the pulsar radiation separating into two modes that propagate along different rays paths and have nearly equal intensities.

Our interpretation of the annulus in Figure~\ref{figES} may be compared with that suggested by Edwards \& Stappers \shortcite{ES04}, who interpreted it in terms of generalized Faraday rotation (GFR). The GFR interpretation requires that polarized radiation be incident on a region in which the polarizations of the natural modes is different from that of the incident radiation. Different amounts of GFR in different pulses can then lead to a distribution of points around an annulus, with the annulus oriented orthogonal to the diagonal through the Poincar\'e sphere defined by the polarizations of the natural modes. In one sense the model based on GFR is complementary to the empirical model proposed here. The empirical model is based on probability distributions, and no specific physical assumptions are made in writing down the probability distributions. GFR provides a physical basis for a probability distribution that, according to Edwards \& Stappers \shortcite{ES04}, reflects the distribution of points seen in Figure~\ref{figES}. However, this is not the interpretation that we suggest. Our interpretation is based on the polarizations of the two orthogonal modes nearly canceling: provided the two modes have nearly equal intensities their sum is weakly polarized, with a wide spread in polarization. With our assumption that the OPMs are 100\% polarized, if the intensities are markedly different (in a given pulse) then the polarization is necessarily close to that of the mode with the higher intensity. An implication of the interpretation we propose is that the points around the annulus must be weakly polarized. There is no such implication with the GFR interpretation.

\subsection{Separation into two modes}

There is a major difficulty in understanding how radiation can be produced in two modes with nearly equal intensities and significantly different ray paths.

How the radiation becomes separated into two natural modes is poorly understood. Most radio emission mechanisms favor radiation into a single natural mode. For example, any maser process that causes one mode to grow faster than the other leads,  after many growth times, to the faster growing mode completely dominating. In principle, this need not be the case: if the growth rate is larger than the rate of generalized Faraday rotation, and if the maser is intrinsically polarized with a polarization different from that of either natural mode in the medium, then the growing radiation can be an intrinsic mixture of the two natural modes \cite{MJ04}. Although the conditions required for this to apply seem implausible, the alternatives seem even less plausible. The alternative is that the emission mechanism results in a single mode, and that the separation into two modes occurs somewhere along the propagation path. Although mode coupling does occur due to inhomogeneity, it is usually a weak effect, whereas the interpretation of OPMs requires comparable intensities in the two modes. This is especially the case for the interpretation of the broad spread in polarization shown in Figure~\ref{figES}: our modeling suggests that this implies that $I_2$ is strongly correlated with $I_1$. Effective mode mixing could occur due to reflection of waves off a sharp boundary, which would need to be near the source region to be effective. (Far from the source the refractive indices are very close to unity, precluding significantly different ray paths for the two modes, as seems to be essential for the interpretation of OPMs.) However, there is no model which incorporates reflection off sharp boundaries. Moreover, near the source one of the modes should have wave properties close to the vacuum, referred to as the X~mode by Arons \& Barnard \shortcite{AB86}, and this neither reflects off sharp gradients nor otherwise couples to the other mode. In brief, it is very difficult to account for pulsar radiation that is an approximately equal mixture of two natural modes, despite the overwhelming observational evidence that the radiation is such a mixture. Hopefully, further use of empirical models will lead to information on the polarizations of the OPMs that will help constrain the mechanism that leads to the separation into two modes.

\subsection{Limiting polarization}

The observed polarization is clearly elliptical in some cases, implying that the natural modes of the pulsar plasma are elliptical at the point where the radiation effectively escapes the magnetosphere. As the radiation in a given mode propagates through the medium, its polarization adjusts continuously so that it remains that of the natural mode at every point along the ray path. (A small leakage into the other mode occurs, implying some mode coupling.) This putative point is referred to as the polarization limiting region, which may be defined as the region beyond which the medium becomes ineffective in changing the polarization of the radiation propagating through it. Polarization limiting is most likely to occur near the cyclotron resonance where the polarization of the natural modes is changing fastest as a function of distance along the ray path \cite{ML04}.  Thus the empirical modeling of the polarization provides information on the polarization limiting region, and on how it varies statistically from pulse to pulse. Such information should help to identify the location of the polarization limiting region.
 
\section{Conclusions}

In this paper we present an empirical model that is useful in simulating data on the Stokes parameters for observations of single pulses at a given pulsar phase. The observed polarization changes from pulse to pulse, and the polarization in any given pulse can be quite different from the (mean) polarization found by summing over a large number of pulses. The underlying model for the interpretation of the observed polarization is in terms of two OPMs: the observed radiation is assumed to be a mixture of the two (completely polarized) OPMs, with intensities $I_1$, $I_2$, that are partially but not completely correlated. We model this by assuming that the intensity for mode~1 has log-normal statistics, with the mean intensity set to unity without loss of generality and the variance in the natural log set to $\Delta_1=0.8$ based on observation, \cite{cjd04}. The ratio, $k$, of the intensities in mode~1 and~2 is assumed to have a gaussian distribution with a mean $\Iot$ and a variance $\dfIt$. The results of the simulations are sensitive to this correlation, and the choice of these parameters are severely constrained by the data, with $\Iot$ near but not equal to unity and $\dfIt$ small but nonzero, cf.\ Table~1. 

In the empirical model presented here, we assume that the polarization of each OPM is determined by a probability distribution, with a mean and a variance on the Poincar\'e sphere. The model allows some non-orthogonality in the OPMs. The intensity of mode~1 is assumed to vary from pulse to pulse with log-normal statistics. The intensity of mode~2 is partially but not totally correlated with the intensity of mode~1. When the polarization of each simulated pulse is plotted on the Poincar\'e sphere, pulses with $I_1\gg I_2$ and $I_1\ll I_2$ lead to peaks around the centroids for modes~1 and~2, respectively, and  pulses with $I_1\approx I_2$ lead to a spreaded out distribution of (weakly polarized) points on the Poincar\'e sphere. The model also allows a single pulse to consist of emission from a number, $N$, of subsources. In the case where we have multiple subsources, we choose the statistical parameters (means and variances) of each subsource to be the same. This leads to some difficulties that we overcome by a somewhat artificial assumption that allows both concentrations around the OPMs and a broad spread of weakly polarized points, as required to simulate some observations. An alternative (also artificial) assumption would be to allow the statistical parameters to be different for different subsources, relying on some subsources to produce the broad spread and other subsources to produce the well-formed peaks about one or both OPMs. Other correlations in the simulations are shown through histograms for the four Stokes parameters, and also for the position angle and the degrees of linear and circular polarization. Such histograms make it possible to describe the variations in the intensity and in the degree of polarization, neither of which are represented on the Poincar\'e sphere. 

Our model contains a large number of free parameters, with this number increasing approximately as the power of the number of subsources. However, given the general framework of the OPM model, this number of parameters is needed to describe a single subsource, and one must rely on comparison with observation to reduce or otherwise constrain the number of free parameters. 

The observational examples on whose interpretation we concentrate, cf.\ Figure~\ref{figES}, includes one phase with well-defined OPMs and another with a concentration of points around one OPM and a diffuse distribution of points that extends around an annulus passing only roughly through the antipodean point on the Poincar\'e sphere. The spread-out case is the more difficult to explain. We find that it requires non-orthogonal OPMs with tightly correlated intensities, with best fits found for the mean and variance of the ratio, $k=I_2/I_1$, of the intensities having the values $\Iot=1.1$, $\dfIt=0.05$. However, this precludes a well-formed peak around either mode, and an additional assumption is needed to account for both a concentrated peak for one mode and a broad spread for the other. We show that this can be achieved by assuming that some fraction of the pulses contain only one mode. (A physical justification for this assumption is that the other mode is sometimes refracted into a direction that does not intersect the Earth.) The important parameter that distinguishes between the top and bottom panel in Figure~\ref{figES} is the variance in the ratio of the intensities: for $\dfIt=0.5$ the model gives two well-defined OPMs. The explanation is straightforward: a large variance in $I_2/I_1$ implies that most pulses have either $I_2\ll I_1$ or $I_2\gg I_1$, giving a polarization distribution close to that for mode~1 and mode~2, respectively. This further emphasizes the importance of the tight correlation required to account for a broad distribution on the Poincar\'e sphere. Note the implication that the degree of polarization of the points that define the broad distribution must be low (very nearly equal mixtures of two OPMs). However, we should emphasize that although these qualitative conclusions are robust, we do not claim that any specific parameter fit is unique. Indeed, it seems likely that there is considerable freedom in the interplay between the variance of the polarization of one OPM ($a_2,b_2$) and in the ratio the intensities ($\Iot,\dfIt$) in accounting for mode~2 around an annulus, in modeling the lower panel in Figure~\ref{figES} for example. 

Our ultimate objective is to understand the origin of the polarization of pulsar radio emission, and hopefully to use it to constrain pulsar models. Our results provide strong support for a model in which the radiation escapes in two natural waves modes that can be significantly elliptically polarized. Moreover, the intensities of the two modes are tightly correlated in at least some examples that we have considered. We note that there is no accepted emission mechanism that can produce roughly equal intensities in two orthogonal modes, and no known propagation effect that can be effective in producing nearly equal intensities in two orthogonal modes for radiation initially in one mode. Further ideas on how this dilemma might be resolved are needed. 

\section*{Acknowledgment}

We thank Simon Johnston for constructive criticism and Steve Ord for comments on the manuscript. We thank the anonymous referee for insightful comments, some of which have been included in our interpretations.

\end{document}